\documentclass{PoS}
\usepackage{epsfig}
\usepackage{amsmath}
\newcommand{\be}{\begin{equation}}
\newcommand{\ee}{\end{equation}}
\newcommand{\bea}{\begin{eqnarray}}
\newcommand{\eea}{\end{eqnarray}}

\PoS{PoS(LAT2005)023}
\title{The volume dependence of spectral weights 
 and the pentaquark state}
\ShortTitle{The volume dependence of spectral weights 
 and the pentaquark state}
\author{\speaker{Constantia Alexandrou}\\
Department of Physics, University of Cyprus, CY-1678 Nicosia, Cyprus\\
E-mail:\email{alexand@ucy.ac.cy}}
\author{Antonios Tsapalis\\
        University of Athens, Institute of Accelerating Systems
	and Applications, Athens, Greece \\        
        E-mail: \email{tsapalis@cc.uoa.gr}}
\abstract{
Before studying the pentaquark system 
 we examine
the spectral weights of the two lowest scattering states 
in the two-pion system 
in the isospin I=2 channel 
on lattices of size $16^3\times 32$, 
$24^3 \times 32$ and $32^3 \times 64$ at $\beta=6.0$ in
the quenched theory. We find that the  spectral weights scale with the volume  
for large time separations. Therefore
 very accurate data are necessary in order that the  spectral weights
determined on different volumes  yield a ratio 
that is precise enough  to distinguish a scattering
state from a single particle state.
The pentaquark system is studied on the same lattices and scaling of the spectral weights of
the low lying state is investigated. The accuracy of
the results obtained for the scaling of spectral weights do not allow us to 
exclude a pentaquark resonance.}

\FullConference{XXIIIrd International Symposium on Lattice Field Theory\\
                 25-30 July 2005\\
                 Trinity College, Dublin, Ireland}

\begin{document}
\section{Introduction}
Several experiments performed during 
the past couple of years (see e.g.~\cite{negative}) 
to search for the $\Theta^+$,
 an exotic baryon  with 
an unusually narrow width, failed to confirm 
the signal that was originally reported
in various  low energy experiments~\cite{positive}.
 This has  cast doubts on the existence of this state 
and raised interesting questions regarding its production mechanism
that add to the puzzle  as to what its structure might be 
to explain its narrow width.

During the same time several quenched lattice calculations   
reached different conclusions as to the existence 
of a resonant five quark state. 
The main difficulty comes from the fact that the  $\Theta^+$ is not 
the lowest lying state but it is expected to be about $100$~MeV above the KN threshold.
 Identifying in 
an unambiguous way the resonance from the KN s-wave 
scattering state within lattice QCD, 
 given the small energy gap, 
is a  difficult task.  One approach employed in a number 
of studies  to distinguish them is to  examine the volume dependence of local 
correlators~\cite{Mathur}. 
Expanding the  correlator computed on a lattice of spatial size $L$ in terms of  
eigenstates of the theory with the same quantum numbers as the interpolating field
 one obtains~\cite{Luscher}:
\be
C_L(t)=\sum_{n=1}^{\infty} w_L^n e^{-E_L^n t} \quad .
\ee
For a single particle state the spectral weights $w_L^n$ 
are approximately volume independent, whereas for a two particle scattering 
state well below resonance $w_L^n \sim 1/L^3$.
Besides the spectral weights one can examine the volume dependence of the energy spectrum. 
For two non-interacting 
particles $h_1$ and $h_2$ in the center of mass frame the energy is given by:
\be
E^n_{h_1 h_2}=\sqrt{m_{h_1}^2 + n\left(\frac{2\pi}{L}\right)^2}+\sqrt{m_{h_2}^2 + n\left(\frac{2\pi}{L}\right)^2}\qquad n=0,1,2,...
\ee
For $n>0$ the energy is volume dependent and it can be distinguished from the energy of a resonance,
 which is volume independent for large enough volumes. 
Before examining the pentaquark system we study the scaling of spectral weights
for a simpler system of two pions  in the isospin $I=2$ channel 
where we expect no low lying resonance.
We use lattices of size $16^3\times32$, $24^3\times32$ and $32^3\times64$ at $\beta=6.0$
with Dirichlet boundary conditions in the temporal direction. 
For a two particle scattering state the expected ratio of spectral weights for our three volumes
is $w_{16}/w_{24}=3.4$, $w_{24}/w_{32}=2.4$ and $w_{16}/w_{32}=8$.
All the results shown here are done 
taking $\kappa=0.153$ 
for the u- and d- quark propagators corresponding to a pion mass of about 830~MeV.

\section{Two pion system}

We  use  local $I=2$ interpolating fields constructed by taking products of pion and rho fields:
\be
J_1(x)=J_1^\pi (x) J_1^\pi (x)\quad J_2(x)=J_2^\pi (x) J_2^\pi (x)\quad
 J_3(x)=J_0^\rho(x) J_0^\rho(x)\quad
 J_4(x)=\sum_{i=1}^3J_i^\rho(x) J_i^\rho(x) 
\ee
where
$
J_1^\pi (x)=\bar{d}(x)\gamma_5u(x),  J_2^\pi (x)=\bar{d}(x)\gamma_5\gamma_0 u(x), 
J_0^\rho(x)=\bar{d}(x)\gamma_0\sum_{i=1}^3\gamma_i  u(x) $ and
$J_i^\rho(x)=\bar{d}(x) \gamma_i  u(x)
$.
\begin{figure}[h]
\begin{minipage}[t]{0.32\linewidth}
\begin{center}
\epsfig{file=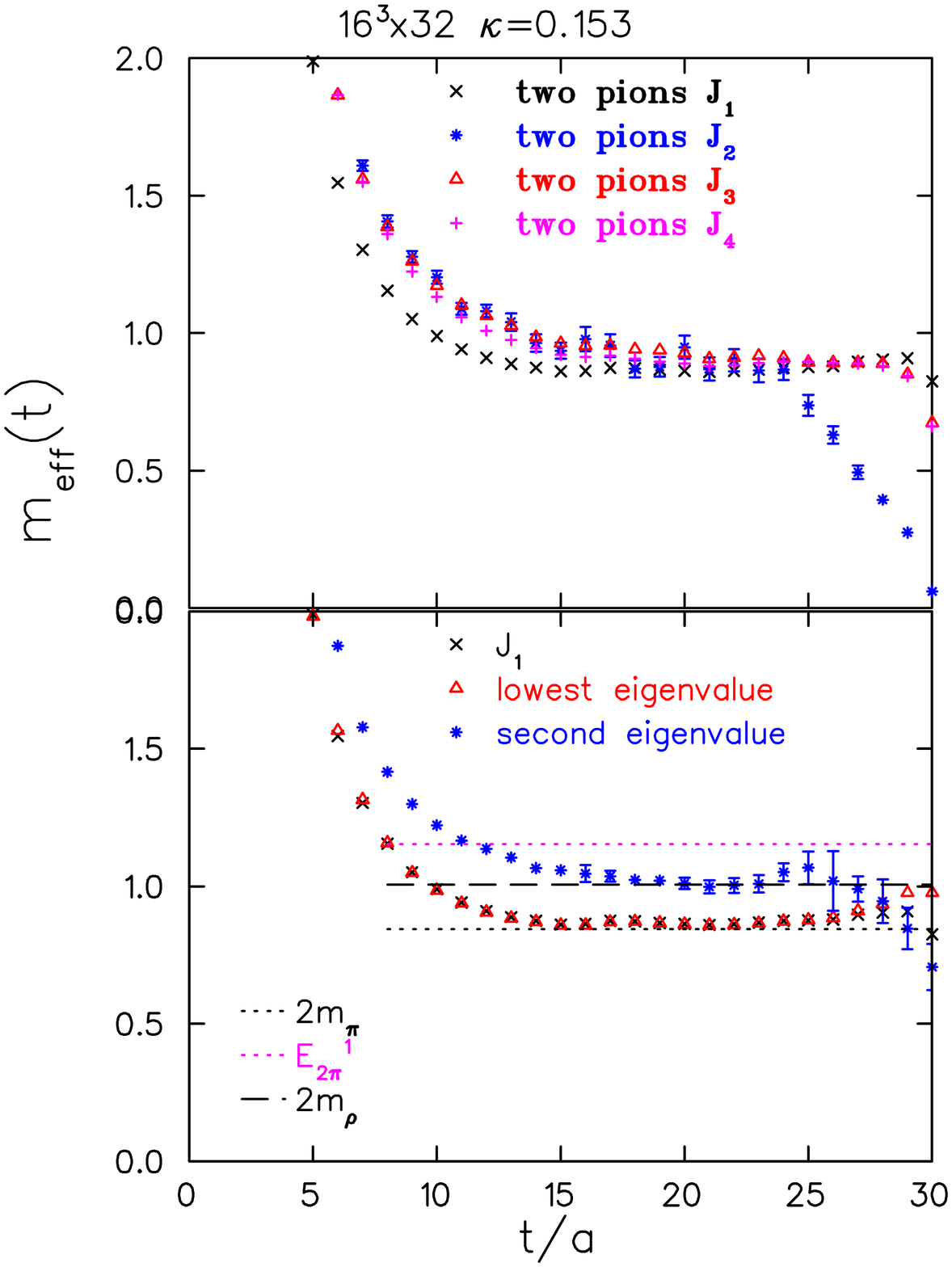,scale=0.25}
\end{center}
\end{minipage}
\begin{minipage}[t]{0.32\linewidth}
\begin{center}
\epsfig{file=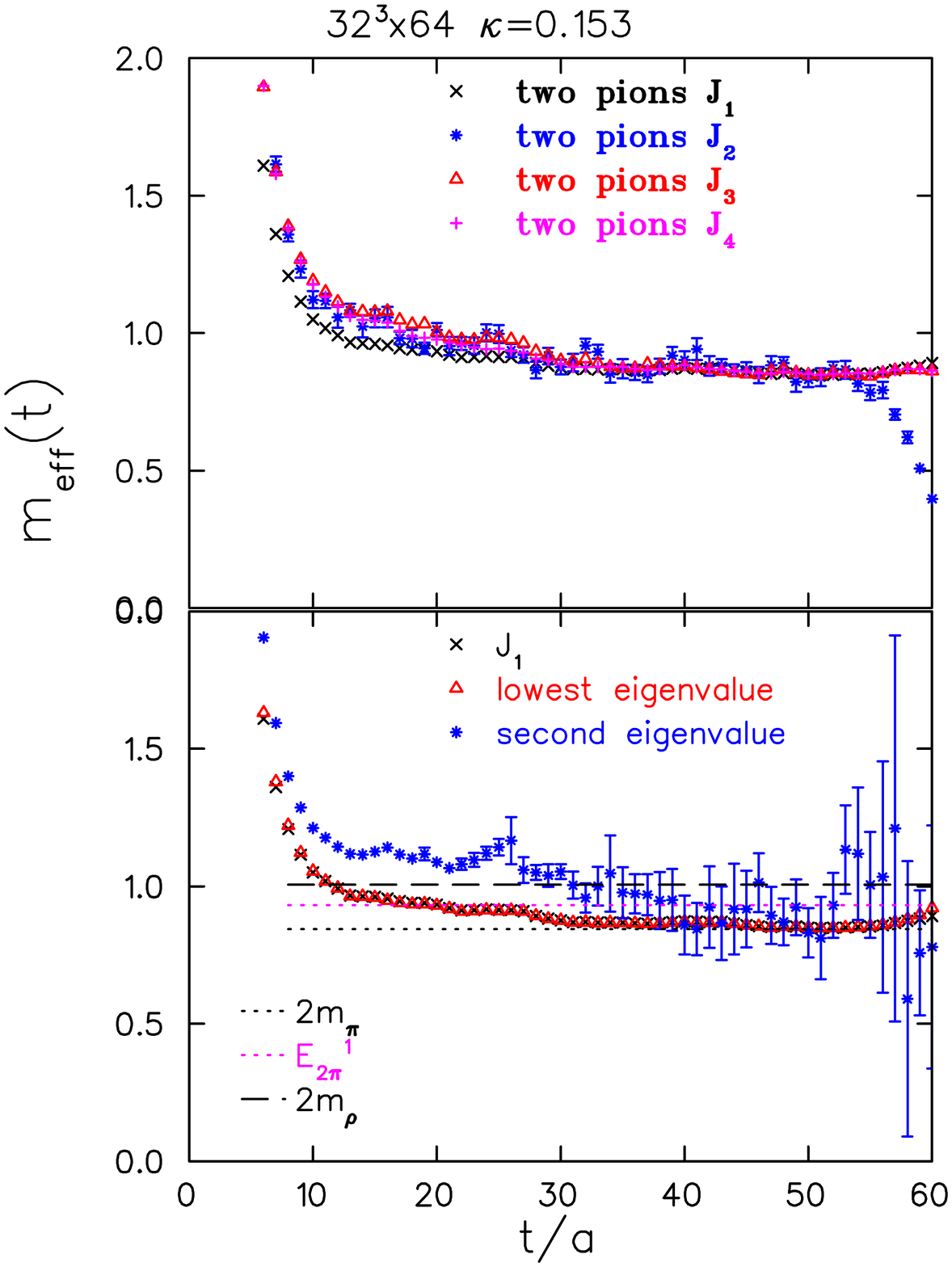,scale=0.25}
\end{center}
\end{minipage}
\begin{minipage}[t]{0.32\linewidth}
\begin{center}
\epsfig{file=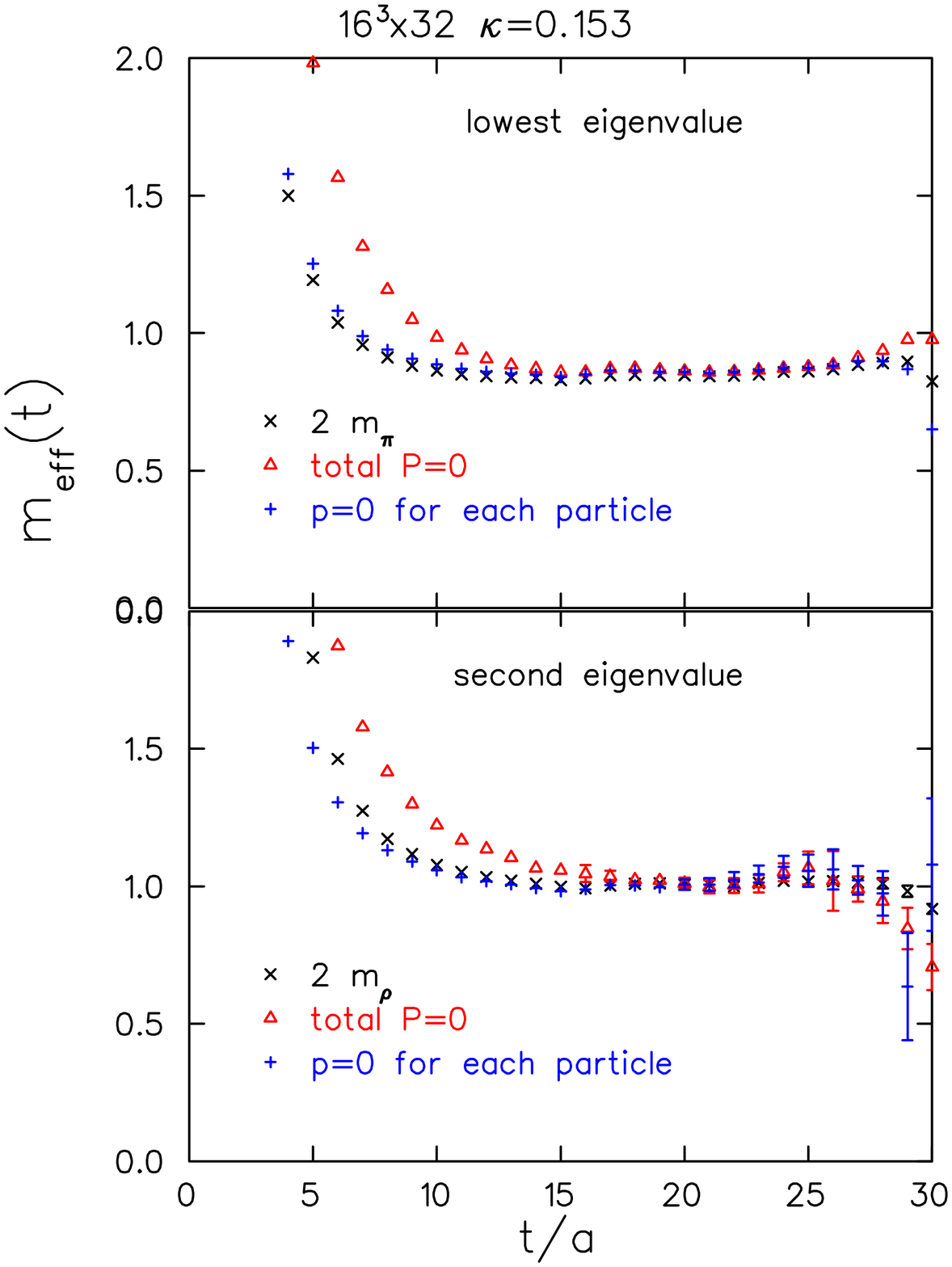,scale=0.25}
\end{center}
\end{minipage}
\hfill
\caption{Upper left and middle graphs show 
effective masses  using 
 $J_1$, $J_2$, $J_3$ and $J_4$
 on a lattice of size $16^3\times32$ and  $32^3\times64$ respectively.
Lower left and middle graphs show effective masses 
 from $J_1$ and   the two lowest eigenvalues.
The graph on the right compares the effective mass for the two lowest eigenvalues
using the unprojected correlation matrix
and the projected correlation matrix of Eq.~(2.2) for the lattice of size $16^3\times 32$.} 
\label{fig:meff pion}
\vspace*{-0.3cm}
\end{figure}
Using the fields defined in Eq.~(2.1) we construct a $4\times4$ correlation matrix 
 $C_{jk}(t)=\sum_{x}\left<0\right|J_j(x)J_k^+(0)\left|0\right>$. In addition we 
project to a state in 
which each of the two particles carries zero  relative momentum by evaluating 
the correlation matrix:
\be
C_{j_{s^\prime}k_s}(t)=\sum_{x,y}\left<0\right|J^{s^\prime}_j(x)J^{s^\prime}_j(y)J_k^{s+}(0)J_k^{s+}(0)\left|0\right>\qquad s,s^\prime=\pi,~\rho
\label{proj c}
\ee
Our variational methods of analysis
are described in Ref.~\cite{paper}.

\begin{figure}[h]
\begin{minipage}{0.5\linewidth}
\begin{center}
\epsfig{file=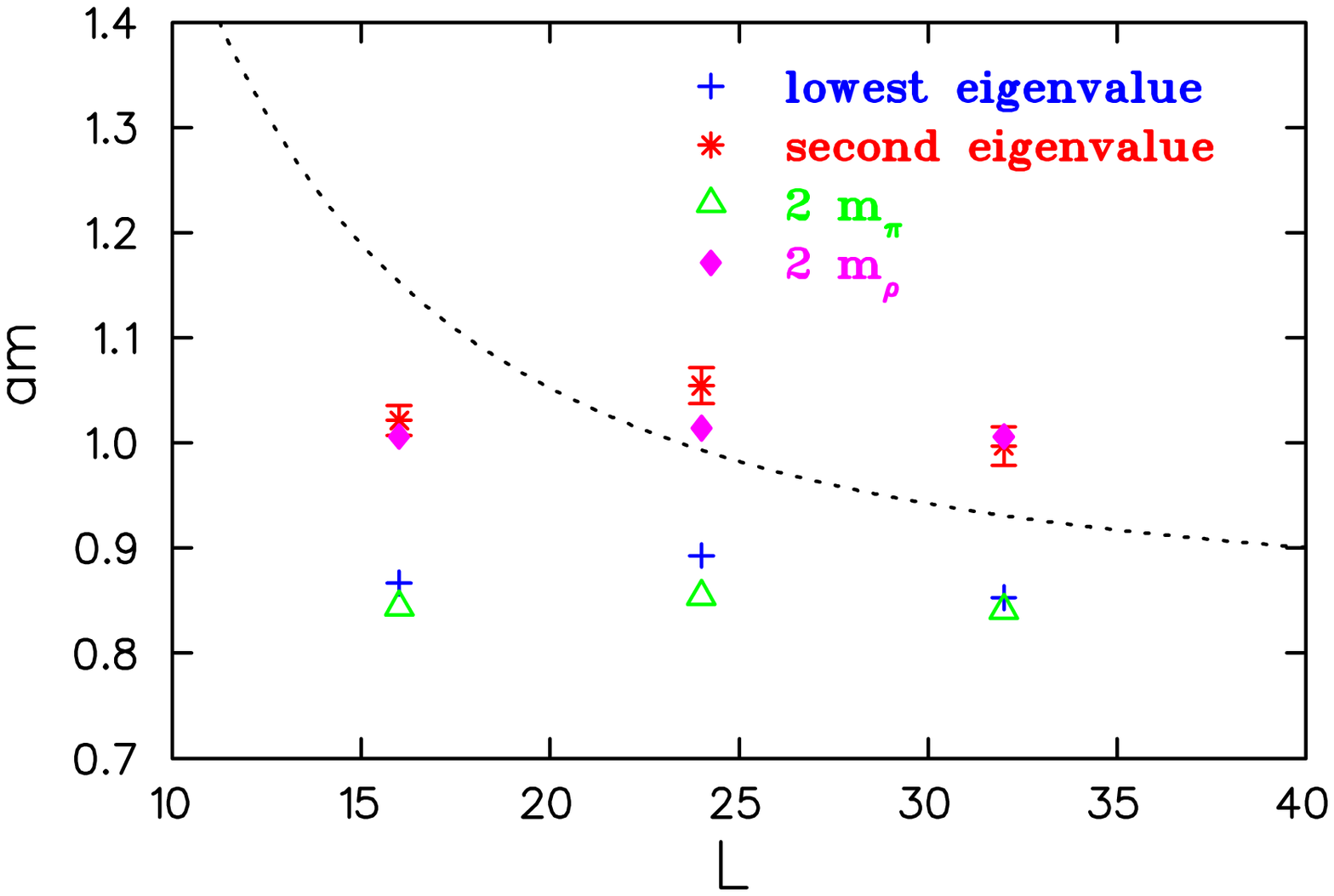,scale=0.35}
\caption{The mass extracted from  the lowest and second lowest energy
 eigenvalue versus L in lattice units.The dotted line is $E_{2\pi}^1$.} 
\label{fig:vol}
\end{center}
\end{minipage}
\hfill
\begin{minipage}{0.45\linewidth}
\begin{center}
\epsfig{file=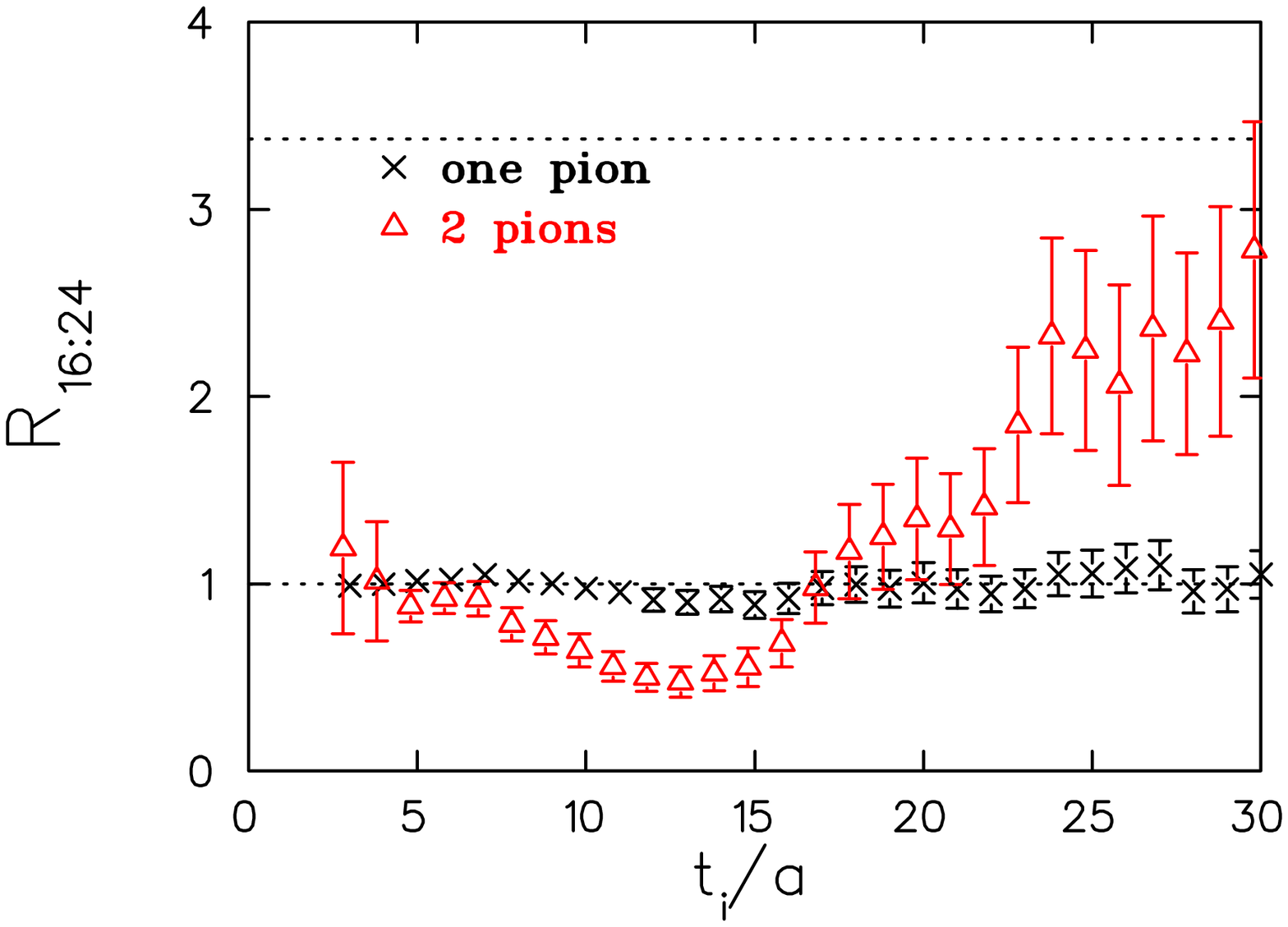,scale=0.33}
\caption{The ratio $R_{16:24}(t)$ as a function of $t$.}
\label{fig:R pion}
\end{center}
\end{minipage}
\end{figure}

The effective masses for  all interpolating fields 
are shown in Fig.~\ref{fig:meff pion} where 
one can see that they converge to the same plateau
 yielding the same value 
for the mass. This value
is the same as  that obtained from the lowest energy eigenvalue 
and  very close to the 
mass of the s-wave two-pion scattering state $E_{2\pi}^0$,
whereas the second eigenvalue yields $E_{2\rho}^0$. 
The fact that the two lowest eigenstates 
correspond to s-wave scattering states can be explicitly demonstrated by 
analyzing the correlation matrix with projection to zero
momentum for each particle 
as shown in the same figure for the lattice of size $16^3\times 32$.
The masses extracted from the two lowest eigenvalues 
are shown 
as a function of the spatial size of the lattice in Fig.~\ref{fig:vol}. 
The energy of the second eigenvalue is volume 
independent and can be clearly distinguished from  $E^1_{2\pi}$.
 Note that finite volume corrections due to particle interactions 
are too small to be seen on the scale of this graph.
The fact that for the lattice of size $24^3\times32$ the masses are slightly 
above   $E_{2\pi}^0$ and  $E_{2\rho}^0$ 
is due to the fact that the time extent of the 
lattice is too small to filter  zero momentum pions and rho mesons.
Also it is worth mentioning  that the two-rho scattering state, 
although  higher than $E_{2\pi}^1$ on our largest lattice,  
is the dominant state in the intermediate time range and
only  for very large times 
the effective mass of the second lowest state becomes consistent with $E_{2\pi}^1$.

\begin{figure}[h]
\begin{minipage}{0.55\linewidth}
\begin{center}
\epsfig{file=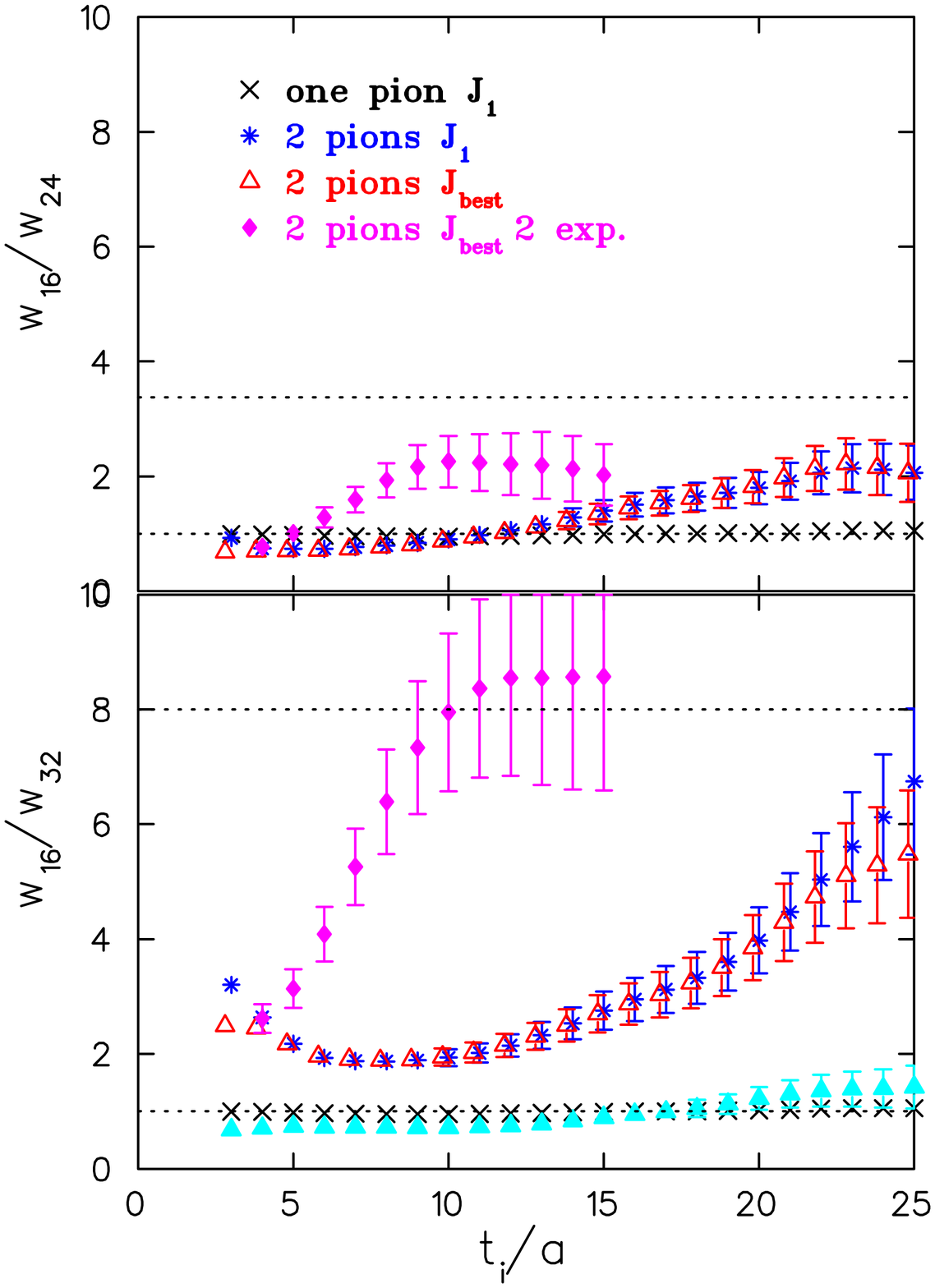,scale=0.35}
\caption{The ratio of spectral weights for the pion and the lowest
state of the two-pion system.
The dotted lines show the expected value of the ratio for a single particle
 and for a two particle scattering
state.
The results shown by the filled triangles
 in  the lower graph are obtained taking the upper fit range for
the large lattice to be  $26$ in lattice units. }
\label{fig:pion weights}
\end{center}
\end{minipage}
\hfill
\begin{minipage}{0.4\linewidth}
\begin{center}
\epsfig{file=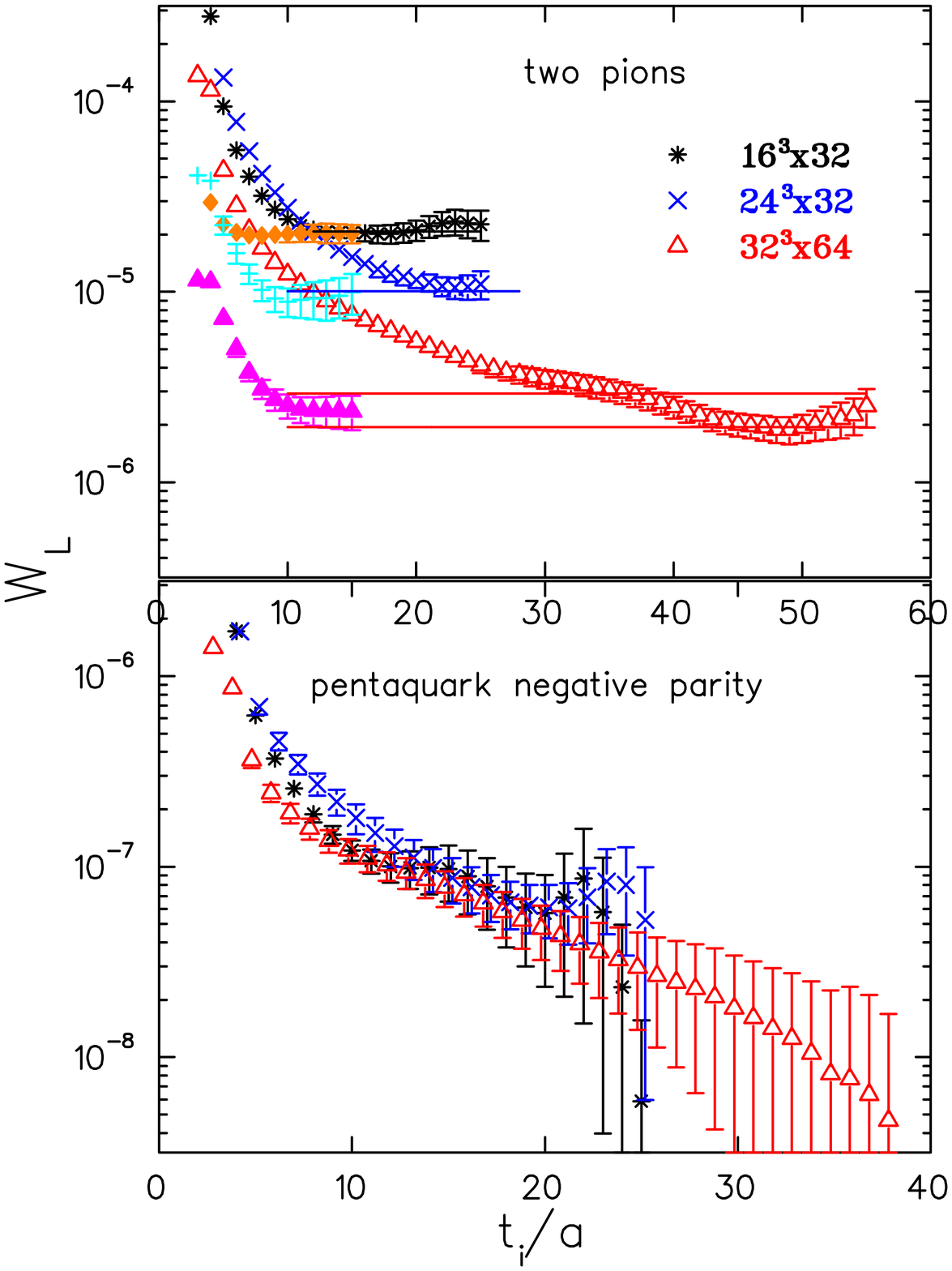,scale=0.35}
\caption{Spectral weights on our three lattices. Top for the two pion system and bottom for
the pentaquark in the negative parity channel. In the pion system
we show results from fits to a single exponential and a sum of two exponentials.}
\label{fig:pion pentaq weights}
\end{center}
\end{minipage}
\vspace*{-0.5cm}
\end{figure}

In order to study  the scaling of spectral weights $w_L$ we first look at
the ratio of correlators computed on lattices of spatial size $L_1$ and $L_2$ 
multiplied by the corresponding effective masses:
\be
R_{L_1:L_2}=\frac{C_{L_1}(t)e^{m_{eff,L_1}(t)}}{C_{L_2}(t)e^{m_{eff,L_2}(t)}} \quad.
\ee
If a single state dominates then this ratio gives the ratio of spectral weights. 
We evaluate this ratio for the lattices of spatial extension $16$ and $24$. 
As can be seen in Fig.~\ref{fig:R pion}, for the one pion state  
this ratio is one as expected. 
For the two-pion state it increases 
but only approaches the expected value of $3.4$ for $t/a>25$ when
the ground state dominates. 
For the lattice of spatial size $32$ this happens for $t/a>30$ 
and therefore this ratio is about one up to $t/a=30$, which is the maximum
time separation that it can be constructed.
A second option is to extract the spectral weights by fitting the correlators 
to one or a sum of two exponentials. 
This allows to take into account information from 
the full time extent of the lattice. 
 We plot the ratio of spectral weights $w_{L_1}/w_{L_2}$ for our three
lattices as a function of the lower time range  $t_i/a$ used in the  fit. 
The upper time range is fixed to $26$ for the lattices of time extent $32$ and to $56$ for
the lattice of time extent $64$. 
As can be seen in Fig.~\ref{fig:pion weights} 
the ratio of weights deviates from one and approaches the expected ratio
 from below for large values of $t_i/a$.
The individual spectral
weights are shown in Fig.~\ref{fig:pion pentaq weights} and they are
volume dependent. 
The values  obtained by
fitting the correlator to a sum of two exponentials 
are consistent with those obtained using a single exponential but converge
at smaller values of $t_i/a$ on all lattices. 
However if instead of an upper fit range of 56 
we take $26$ for the $32^3\times64$ lattice 
as we did for the other two then the ratio of spectral weights
stays very close to one as can be seen in Fig.~\ref{fig:pion weights}. 
The same analysis can be done for the second eigenstate but the errors are too
large to reach a definite conclusion even for this simple system.

\section{Implications for the pentaquark system}
We use interpolating fields motivated by the diquark-diquark~\cite{sasaki}
and KN structure: 
\be
J_{DD}^{I=0}=\epsilon^{abc}\left(u_a^TC\gamma_5 d_b\right)
\left[u_c^TCd_e- u_e^TCd_c \right] C\bar{s}_e^T\gamma_5\>,\quad 
J_{KN}^{I=0}=\epsilon^{abc}\left(u_a^TC\gamma_5 d_b\right)\biggl[u_c\left(\bar{s}\gamma_5d\right)- d_c\left(\bar{s}\gamma_5 u\right)\biggr].
\ee
\begin{figure}[h]
\begin{minipage}[t]{0.53\linewidth}
\begin{center}
\epsfig{file=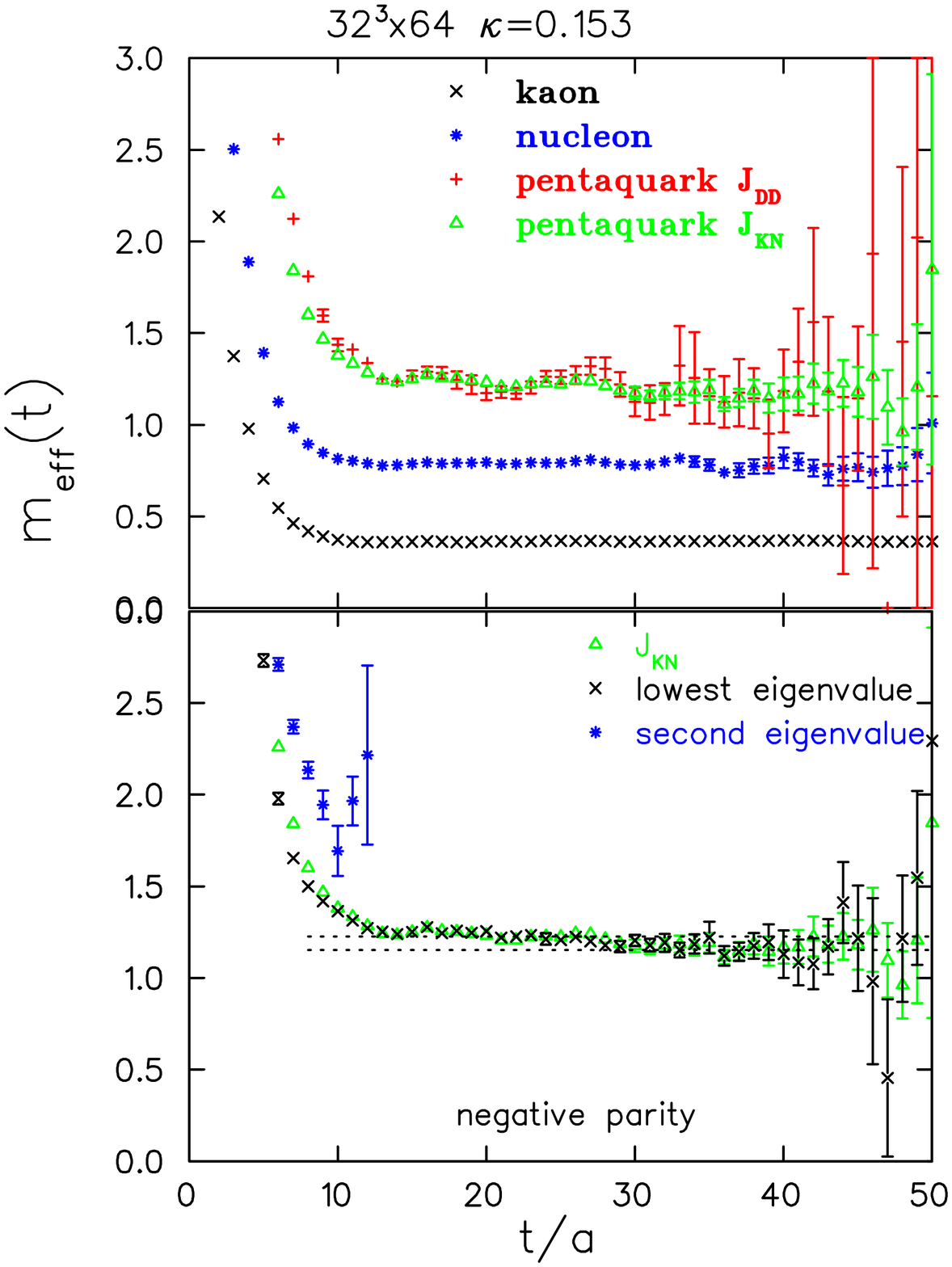,scale=0.3}
\caption{The effective mass for the pentaquark in the negative parity channel.
Top: for $J_{DD}$ and $J_{KN}$. Bottom: for the two lowest eigenvalues.
 The dotted lines show the energies of the KN scattering 
states $E_{KN}^0$ and $E_{KN}^1$.}
\label{fig:pentaq negative}
\end{center}
\end{minipage}
\hfill
\begin{minipage}[t]{0.4\linewidth}
\begin{center}
\epsfig{file=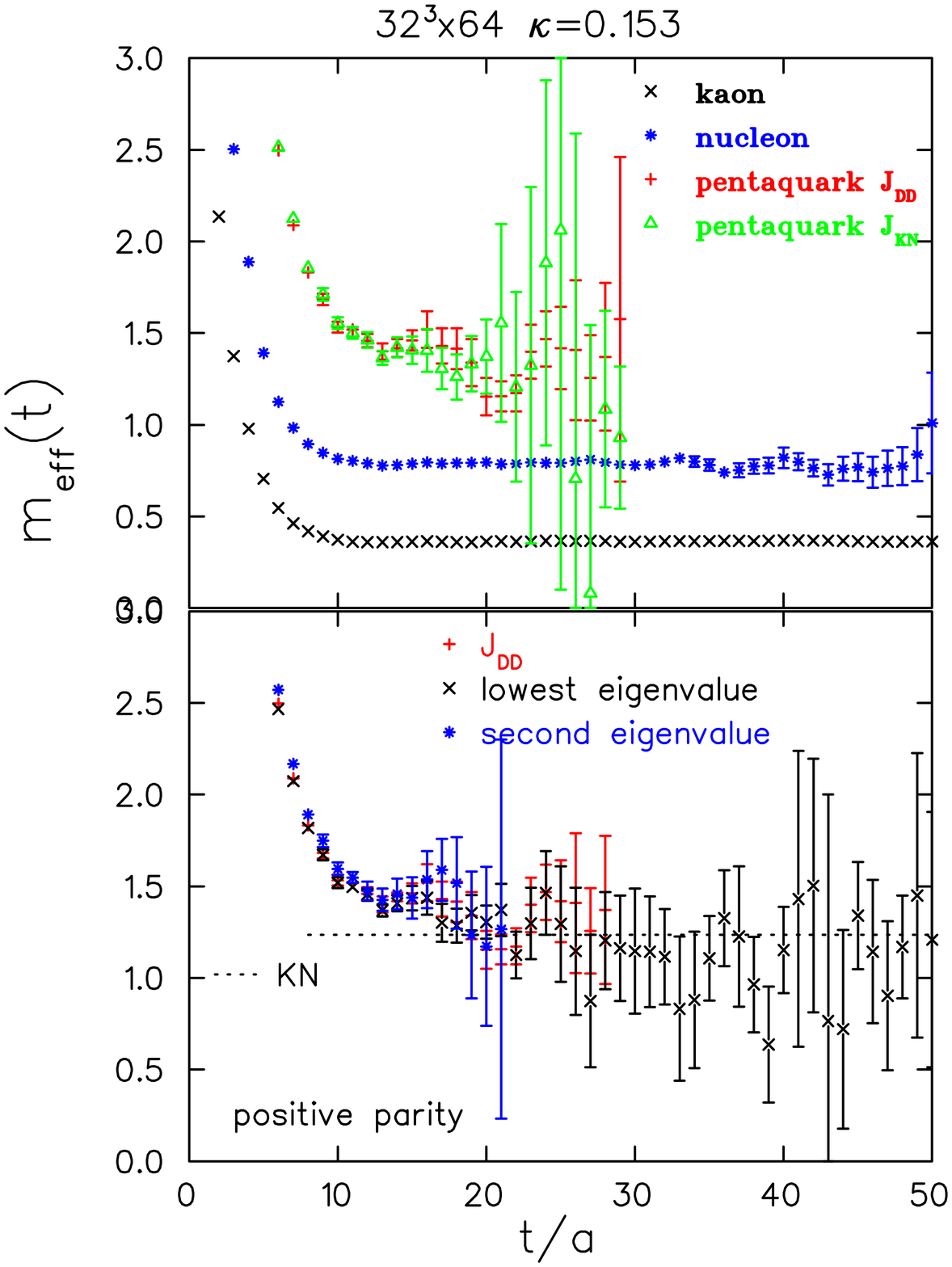,scale=0.3}
\caption{The same as Fig.~8 put for the positive parity channel.
 The dotted line is $E_{KN}^1$.}
\label{fig:pentaq positive}
\end{center}
\end{minipage}
\end{figure}
We fix  $\kappa_s=0.155$ for the strange quark. This choice gives  $m_K/m_N=0.5$ 
and $m_\phi/m_N=1.04$ 
close to the experimental ratios.
The effective masses 
for the negative and positive parity channels are shown in Figs.~\ref{fig:pentaq negative}
 and \ref{fig:pentaq positive} 
for the 
$16^3\times32$ and $32^3\times 64$ lattices.
 The plateau region is larger for the negative parity than 
for the positive parity channel with both interpolating fields yielding 
consistent results. For comparison we also show the effective mass for the nucleon and the kaon.
We perform the same variational analysis as in the two pion system to extract 
the two lowest  energy eigenvalues shown 
in Figs.~\ref{fig:pentaq negative} and \ref{fig:pentaq positive}.
What we find is that in the negative parity channel the second energy is very poorly determined 
whereas the lowest state is consistent with that 
obtained from either interpolating fields. Like in the two pion system the s-wave KN scattering 
state is obtained when $t/a>30$ for the large lattice. 
For the positive parity the two eigenvalues are very close together 
and cannot be accurately resolved.
We find a  mass splitting of roughly $100$~MeV at this quark mass.
As in the two pion system only the  ratio $R_{16:24}$ is useful. 
For the $32^3\times64$ lattice the ratio is not useful 
since one has to go beyond $t/a>30$ to see the scaling.
Instead we extract  the spectral weights by fitting 
the correlators to a single exponential
using  the same upper fit ranges as those used in the two-pion system. 
The results are shown in Fig.~\ref{fig:pion pentaq weights}.
 In both the two-pion and pentaquark systems the values of the spectral weights
 on the two smaller lattices stabilize  
as we increase  the lower fit range $t_i/a$  whereas 
on the large lattice convergence is slow. 
 However  within
this variation the ratio of weights clearly deviates from unity.
For the pentaquark system
fitting to two exponentials is very noisy and only single exponential fits
are performed.
 The values of the weights extracted on the 
three lattices show a very different
behavior as compared to that observed in the two-pion system.
Within the statistical errors they show no volume
dependence  for $10<t_i/a<20$ unlike in the two pion system where
for the same time range  a clear
volume dependence is seen.

\section{Conclusions}

The mass correlation matrix constructed from operators that
have the quantum numbers of the isospin I=2 two pion system 
yields two eigenvalues that 
 correspond to the two-pion and two-rho s-wave scattering states. 
Scaling of the spectral weights  with the volume is verified but
requires large time separations
and accurate data.  
Even though for this system the statistical errors 
on the correlators for the first 
excited state are small  the ratio of spectral  weights for this state is  
still too noisy obscuring the volume dependence.
Carrying the same analysis for the pentaquark system we find that,
using the KN and the diquark-diquark  interpolating fields, 
we obtain reliably only
 the lowest energy eigenvalue in the negative parity channel which,  for large time separations,
is consistent with 
$E_{KN}^0$.
 The spectral weights can only be accurately determined
  in the negative parity channel 
and for our three volumes they do not show strong volume 
dependence 
unlike what is observed in the pion system.   
Thus, to  the accuracy with which the scaling of
 the spectral weights is determined,
 we cannot exclude
a pentaquark resonance.
In the positive parity channel  we find that the two eigenvalues 
are very close in energy with  a gap of about
100~MeV at $\kappa=0.153$. However the mass of this state is too high to be
identified as the $\Theta^+(1540)$.


\end{document}